\documentclass[aps,prb,reprint,superscriptaddress]{revtex4-2}
\usepackage[colorlinks=true, linkcolor=blue, citecolor=blue,urlcolor=blue]{hyperref}

\usepackage{svg}
\usepackage{graphicx}
\usepackage{amsmath}
\usepackage{amssymb}
\usepackage{color}
\usepackage{float}

\newcommand{\SIQSE}{\affiliation{1}{Shenzhen Institute for Quantum Science and Engineering, Southern University of Science and Technology, Shenzhen 518055, China}}

\newcommand{\IQA}{\affiliation{3}{International Quantum Academy, Shenzhen 518048, China}}
\newcommand{\GDKL}{\affiliation{4}{Guangdong Provincial Key Laboratory of Quantum Science and Engineering, Southern University of Science and Technology, Shenzhen 518055, China}}
\newcommand{\HFNL}{\affiliation{5}{
Shenzhen Branch, Hefei National Laboratory, Shenzhen 518048, China}}

\definecolor{cola}{rgb}{0.7,0.1,0.1}
\definecolor{colb}{rgb}{0.6,0.6,0}
\definecolor{colc}{rgb}{0.3,0.7,0}
\definecolor{cold}{rgb}{0,0.35,0.75}
\definecolor{cole}{rgb}{0.63, 0.13, 0.94}
\definecolor{colf}{rgb}{0.0, 0.6, 0.0}

\begin{document}
\title{A C-Band Cryogenic GaAs MMIC Low-Noise Amplifier for Quantum Applications}

\author{Zechen Guo}
\affiliation{\SIQSE}\affiliation{\IQA}\affiliation{\GDKL}

\author{Daxiong Sun}
\affiliation{\SIQSE}\affiliation{\IQA}\affiliation{\GDKL}

\author{Peisheng Huang}
\affiliation{School of Physics and Electronic-Electrical Engineering, Ningxia University, Yinchuan 750021, China}
\affiliation{\IQA}

\author{Xuandong Sun}
\affiliation{\SIQSE}\affiliation{\IQA}\affiliation{\GDKL}

\author{Yuefeng Yuan}
\affiliation{\IQA}

\author{Jiawei Zhang}
\affiliation{\SIQSE}\affiliation{\IQA}\affiliation{\GDKL}

\author{Wenhui Huang}
\affiliation{\SIQSE}\affiliation{\IQA}\affiliation{\GDKL}

\author{Yongqi Liang}
\affiliation{\SIQSE}\affiliation{\IQA}\affiliation{\GDKL}

\author{Jiawei Qiu}
\affiliation{\SIQSE}\affiliation{\IQA}\affiliation{\GDKL}

\author{Jiajian Zhang}
\affiliation{\SIQSE}\affiliation{\IQA}\affiliation{\GDKL}

\author{Ji Chu}
\affiliation{\IQA}

\author{Weijie Guo}
\affiliation{\IQA}

\author{Ji Jiang}
\affiliation{\SIQSE}\affiliation{\IQA}\affiliation{\GDKL}

\author{Jingjing Niu}
\affiliation{\IQA}\affiliation{\HFNL}

\author{Wenhui Ren}
\affiliation{\IQA}

\author{Ziyu Tao}
\affiliation{\IQA}

\author{Xiayu Linpeng}
\email{linpengxiayu@iqasz.cn}
\affiliation{\IQA}

\author{Youpeng Zhong}
\email{zhongyp@sustech.edu.cn}
\affiliation{\SIQSE}\affiliation{\IQA}\affiliation{\GDKL}\affiliation{\HFNL}

\author{Dapeng Yu}
\affiliation{\SIQSE}\affiliation{\IQA}\affiliation{\GDKL}\affiliation{\HFNL}


\date{\today}

\begin{abstract}
Large-scale superconducting quantum computers require massive numbers of high-performance cryogenic low-noise amplifiers (cryo-LNA) for qubit readout. Here we present a C-Band monolithic microwave integrated circuit (MMIC) cryo-LNA for this purpose. This cryo-LNA is based on 150~nm GaAs pseudomorphic high electron mobility transistor (pHEMT) process and implemented with a three-stage cascaded architecture, where the first stage adopts careful impedance match to optimize the noise and return loss. The integration of negative feedback loops adopted in the second and third-stage enhances the overall stability. Moreover, the pHEMT-self bias and current multiplexing circuitry structure facilitate the reduction of power consumption and require only single bias line. Operating at an ambient temperature of 3.6~K and consuming 15~mW, the cryo-LNA demonstrates good performance in the C-band, reaching a 5~K equivalent noise temperature and an average gain of 40~dB. We further benchmark this cryo-LNA with superconducting qubits, achieving an average single-shot dispersive readout fidelity of 98.3\% without assistance from a quantum-limited parametric amplifier. The development of GaAs cryo-LNA diversifies technical support necessary for large-scale quantum applications.
\end{abstract}

\maketitle


\section{INTRODUCTION}
\label{1}
Superconducting qubits are one of the leading candidates for building large scale, fault-tolerant quantum computers~\cite{Acharya2023a,Acharya2024,Gao2024,Cheng2024,Yang_2024,Niu_2023,Niu_2024,Wu_2024}, where the number of qubits integrated on a superconducting processor has surged from a handful to a thousand over the past decade~\cite{Barends2013,Chen2014,Arute2019,Zhu2022,Krinner2019}. Such large-scale quantum processors require massive numbers of high-performance cryo-LNAs for qubit readout. However, although LNAs have found mature applications in various disciplines such as telecommunication and radar systems, designing LNAs for cryogenic operation has remained challenging, partly because of changes in device behavior at cryogenic temperatures. Wideband matching also remains difficult to achieve and circuit stability is hard to maintain under low power consumption at such extreme conditions.

Pseudomorphic high electron mobility transistors (pHEMTs)  have been extensively utilized in LNAs at microwave frequencies. In particular, Indium Phosphide (InP) pHEMTs have been the preferred option for achieving ultra-low noise levels at microwave frequencies, especially under cryogenic temperatures~\cite{Wadefalk2003,Cha2020,Cha2020a,Zeng2024,Cha_2023}.
SiGe BiCMOS LNAs have also been shown to exhibit excellent performance at cryogenic temperatures~\cite{Wu2023,Zou2024,Peng2024}, along with cryo-CMOS modeling for quantum applications~\cite{Tang2022,Xue2023}. 
On the other hand, Gallium Arsenide (GaAs) pHEMTs, as a classic choice for microwave amplifier design, boasts extensive applications and reliable performance in industry. Its performance at cryogenic temperatures has also been validated, including cryogenic behaviors of transistors~\cite{Pospieszalski1988,Khandelwal_2013,Ardizzi2022}, and the implementation of cryo-LNA~\cite{Risacher2003,Shimizu2020,Chiong2013,Jiang_2018,Heinz2024}.
GaAs LNA offers the promise of cost-effectiveness and is well-suited for large-scale manufacturing, making it an attractive option for advancing quantum applications at large scale.

\label{1}
\begin{figure*}[t]
    \centering
    \includegraphics[width=0.8\textwidth]{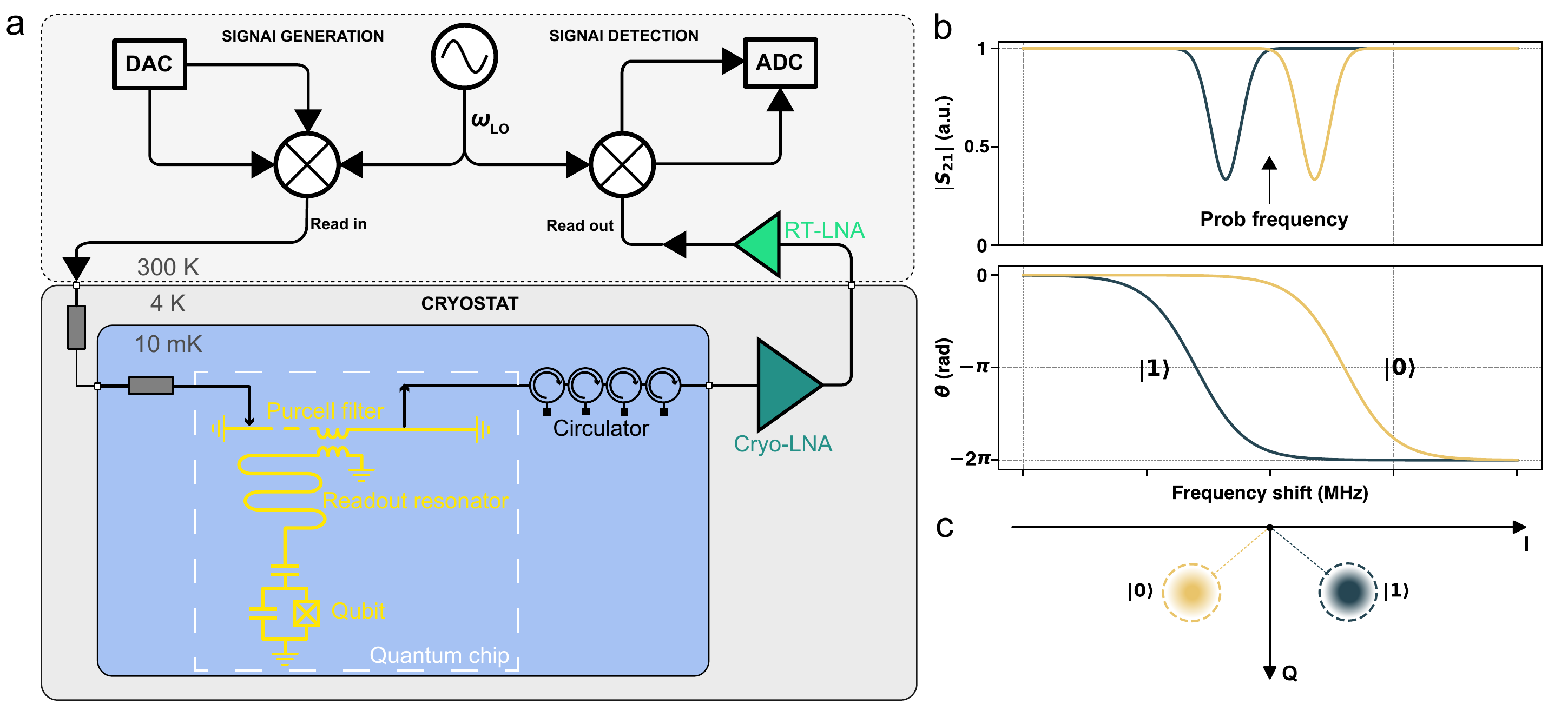}
    \caption{Overview of the readout process for superconducting qubits. (a) Simplified circuit schematic of the superconducting qubit readout system, including both signal generation and detection. The setup includes a signal generation circuit with a local oscillator ($\omega_{\rm{LO}}$), an IQ mixer and a digital-to-analog converter (DAC) for generating the modulated readout tone, and an analog-to-digital converter (ADC) for detecting the readout signal after demodulation by another IQ mixer. (b) Qubit state dependent dispersive shift, illustrated by the transmission magnitude ($|S_{21}|$) and phase shifts ($\theta$) of the prob tone after passing through the readout system, for the ground state ($|0\rangle$) and excited state ($|1\rangle$) respectively. (c) Typical distribution of the integrated readout signal in the I-Q quadrature plane.}
    \label{fig:1}
\end{figure*}

In this paper, we present a C-Band cryogenic GaAs MMIC LNA designed for quantum applications. We first analyze the specifications of the cryo-LNA for superconducting qubit readout. Subsequently, we describe the amplifier's design procedure, including a discussion on the pHEMT current multiplexing structure and the negative feedback loop optimized for cryogenic operation. We then present the device's performance, including transmission and noise measurement results of the cryo-LNA. Finally, we benchmark the cryo-LNA's performance with superconducting qubits, achieving an average single-shot dispersive readout fidelity of 98.3\% without assistance from a quantum-limited parametric amplifier. The advancements in GaAs MMIC cryo-LNA technology diversify the technical support necessary for large-scale quantum applications.

\section{AMPLIFIER SPECIFICATION FOR SUPERCONDUCTING QUBIT READOUT}





To understand the required specifications of the amplifier, we first review the readout process for superconducting qubits~\cite{Krantz2019, Blais2021}. Figure~\ref{fig:1}(a) illustrates a simplified schematic of a typical readout system. The qubit consists of a Josephson junction shunted by a capacitor, forming a non-linear resonator. When cooled to approximately 10~mK in a dilution refrigerator (DR), the energy of the system is confined to its two lowest levels: the ground state $|0\rangle$ and the excited state $|1\rangle$, which constitute the two-level quantum system, or qubit. The qubit is capacitively coupled to a readout resonator in the dispersive regime, causing the frequency of the readout resonator to shift depending on the qubit state~\cite{Blais_2004}, as shown in Fig.~\ref{fig:1}(b). In this regime, a probe tone passing through the resonator acquires a qubit-state-dependent phase shift.

To reduce thermal noise, the probe tone is first attenuated at various cooling stages and then filtered before reaching the readout resonator. An on-chip bandpass Purcell filter, positioned between the readout resonator and the environment, suppresses the Purcell decay rate by blocking microwave propagation at the qubit frequency without affecting the readout speed~\cite{Houck_2008,Jeffrey2014,Walter_2017}. The weak readout signal then passes through a chain of circulators, which act as microwave isolators, before being amplified by cryogenic and room-temperature amplifiers. It is subsequently demodulated by an IQ mixer to extract an integrated signal in the I-Q quadrature, as illustrated in Fig.~\ref{fig:1}(c). Due to noise in the readout chain, the I-Q quadrature signal exhibits a Gaussian distribution, with its phase dependent on the qubit state. By post-processing the distribution to align the centers of the two states along the $Q$ axis, the readout signal-to-noise ratio (SNR) is defined as
\begin{equation} 
\text{SNR}^2 = \frac{(c_1 - c_0)^2}{\sigma_1^2 + \sigma_0^2}, 
\end{equation}
where $c_i$ and $\sigma_i$ represent the mean and standard deviation of the Gaussian distribution for state $|i\rangle$ ($i = 0, 1$)~\cite{Blais2021}. The readout fidelity $F_i$, defined as the probability of correctly assigning the state $|i\rangle$, is related to the SNR by $F_i = 1 - \text{erfc}(\text{SNR}/2)/2$, where $\text{erfc}$ is the complementary error function~\cite{Gambetta_2007}. Achieving high enough readout fidelity requires the cryo-LNA to have a low enough noise temperature $T_e$ (4~K typical) and high enough gain ($>30$~dB) to prevent significant degradation of the SNR during amplification.

\begin{figure*}[!t]
    \centering
    \includegraphics[width=0.9\textwidth]{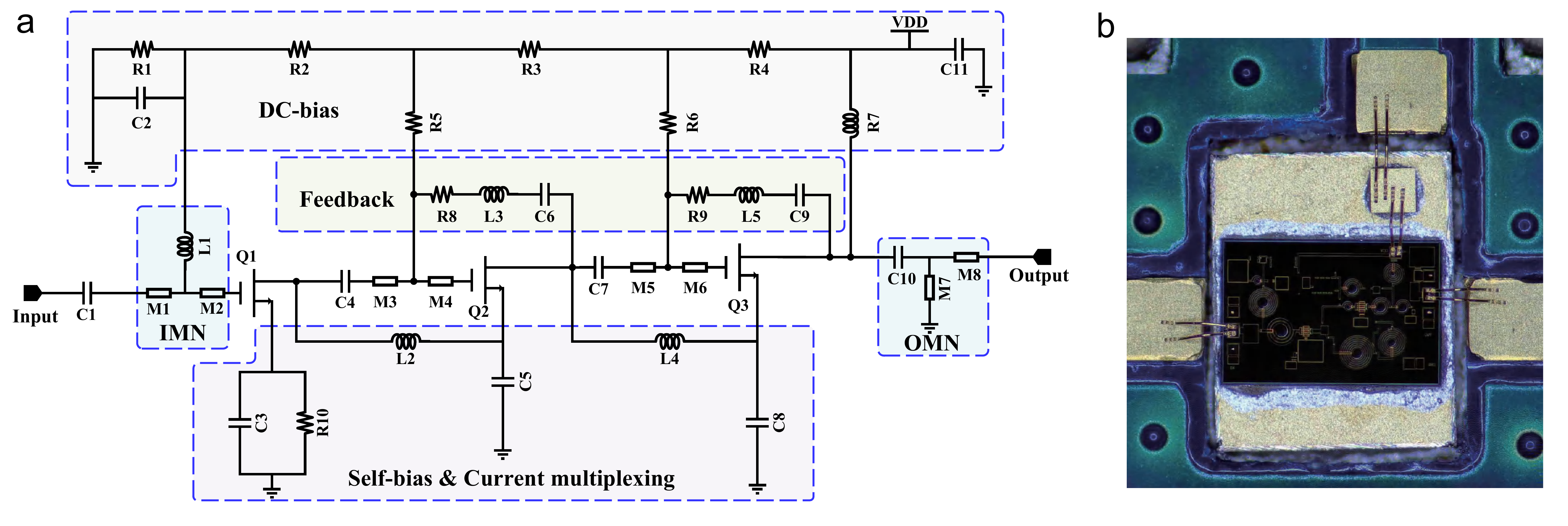}
    \caption{(a) Simplified schematic of the designed cryogenic LNA. (b) A close-up picture of the wire bonded LNA-chip.}
    \label{fig:2}
\end{figure*}

The limited cooling power and space of dilution refrigerators (DRs) impose practical constraints on both the power consumption budget and the size of LNAs. Although the discrete component approach offers greater flexibility during prototype design~\cite{Bardin2021,Peng2023}, the use of MMIC technology enables more compact packaging for scalability and ensures stable performance with low power consumption at cryogenic temperatures. In the context of cryogenic wiring, the fan-out configuration of a single DC bias line within the cryo-LNA becomes essential, particularly as the number of readout channels scales up. However, GaAs pHEMTs typically require two fan-out lines for gate and drain biasing. For circuits incorporating two or more transistor stages, individual bias lines for each stage may result in a fan-out of more than four bias lines, complicating wiring for cryo-LNA applications. Additionally, high current consumption increases the power burden on the amplifier design.
To address these challenges, adopting a self-bias configuration for pHEMTs is desirable. This approach allows the pHEMT to be biased at any percentage of the drain-to-source current ($I_{ds}$) using a single drain bias supply. In terms of the cryo-LNA's power consumption, it is limited by the cooling power of commercial DRs, which is approximately 3~W at the 4~K stage. Considering the prospect of 1000 qubits with 10 qubits per readout line, the individual cryo-LNA's DC power consumption should be constrained to about 10~mW, within the allocated power budget of 1~W for the 4~K stage.



\section{DESIGN AND IMPLEMENTATION}
\label{3}

Figure~\ref{fig:2}(a) depicts the simplified schematic of the GaAs pHEMT LNA in this work. It adopts a three-stage cascaded transistor topology, including a DC bias network, input matching network (IMN), output matching network (OMN), feedback loop, and current multiplexing structure. The LNA is designed using MMIC technology based on a 150~nm GaAs pHEMT process. The first-stage pHEMT biasing condition is initially validated using room-temperature models.

To realize the cryo-LNA operating in the 4~GHz to 8~GHz frequency band, with a gain exceeding 35~dB, $T_e$ below 6~K, and power consumption under 20~mW, a three-stage architecture was chosen as it offers a reasonable trade-off among bandwidth, noise, and power consumption. The matching network of the initial stage was optimized for optimal noise performance~\cite{Sabzi2020}. The second stage was fine-tuned to provide ample gain, and the third stage was engineered to enhance gain flatness while achieving extensive output impedance matching.

Due to the limited cooling power in the DR, a current multiplexing structure was implemented to minimize the power consumption of the designed LNA. In this scheme, an inductor is connected in series from the source to the pre-stage drain. As shown in Fig.~\ref{fig:2}(a), the drain current of the second and third stages flows from their sources to the drain of the pre-stage. The sources of all transistors are AC-grounded through capacitors. For a single-stage transistor biased at $-0.6~\rm{V}/15~\rm{mA}$, the current can be efficiently multiplexed by other transistors, significantly reducing power consumption from $\sim75~\rm{mW}$ to $\sim34~\rm{mW}$ at 300~K compared to a direct power supply scheme.
For cryogenic broadband applications, a negative feedback loop was adopted to achieve flat gain and enhance cryogenic stability. In this design, an RLC series network is connected from the drain to the gate, facilitating input and output impedance matching.

A die photograph of the chip is shown in Fig.~\ref{fig:2}(b), with a compact size of 1.85~mm $\times$ 1.2~mm. The cryo-LNA was packaged in a copper box with 2.92~mm connectors and insulator terminals. The microstrip lines for signal transmission were made of Rogers 5880. The dimension of the LNA assembly is 18~mm×18~mm×9~mm (without connectors).
\begin{figure*}[t]
    \centering
    \includegraphics[width=0.8\textwidth]{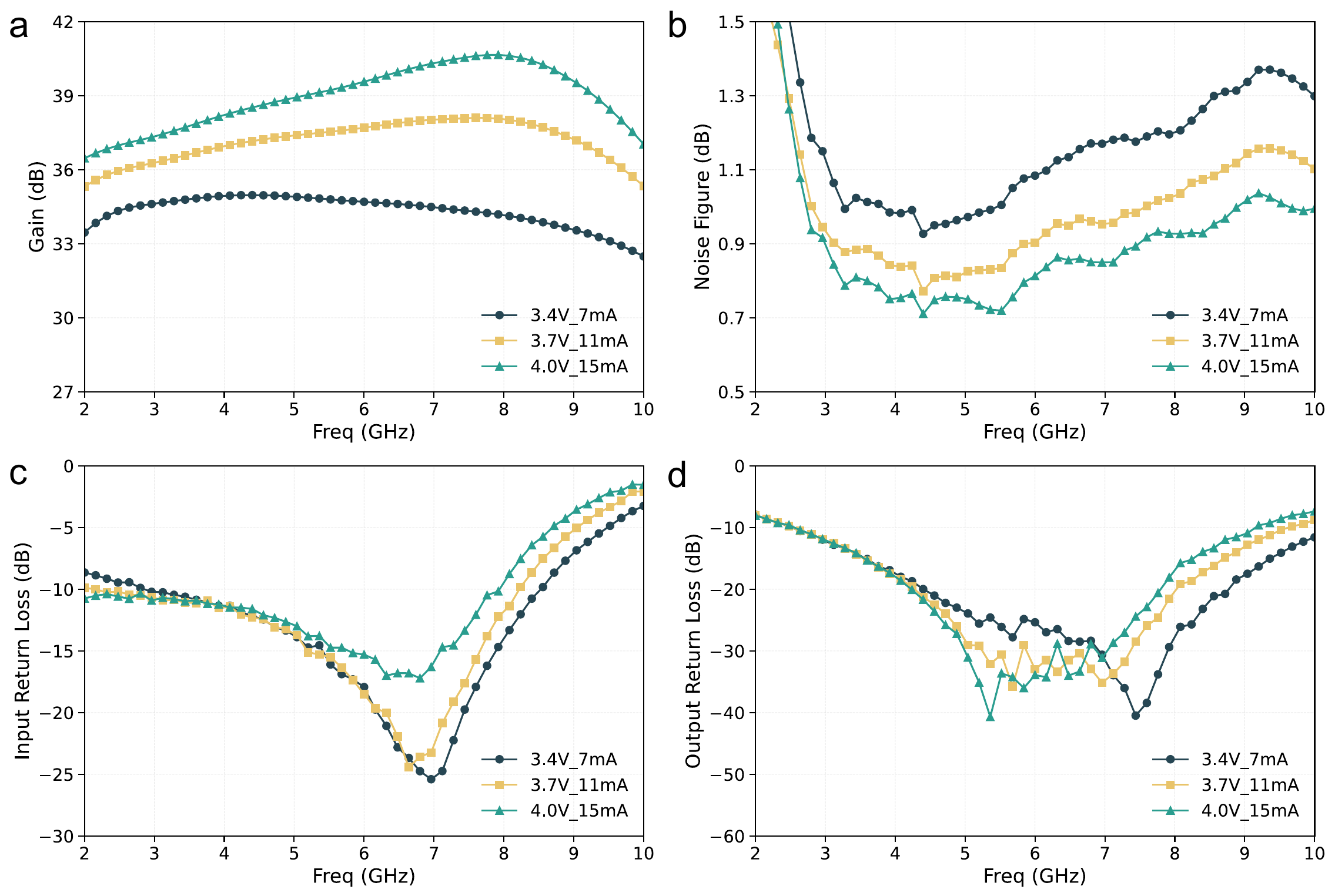}
    \caption{The cryo-LNA performance at ambient temperature ($T_a$ = 300~K). (a) Gain performance across the 2–10~GHz frequency range, showing an increase in gain with higher supply voltage and a maximum gain exceeding 40~dB. (b) Noise figure, remaining below 1.5~dB across the entire frequency range, with a minimum value of less than 0.9~dB in the 5–7~GHz range. (c) Input return loss ($S_{11}$) demonstrating good input impedance matching with values below $-10$~dB in the 4-8~GHz range. (d) Output return loss $S_{22}$ illustrating effective output impedance matching with values below $-15$~dB in the 4-8~GHz range.}
    \label{fig:3}
\end{figure*}

\section{DEVICE PERFORMANCE}
\label{4}

\begin{figure*}[t]
    \centering
    \includegraphics[width=0.8\textwidth]{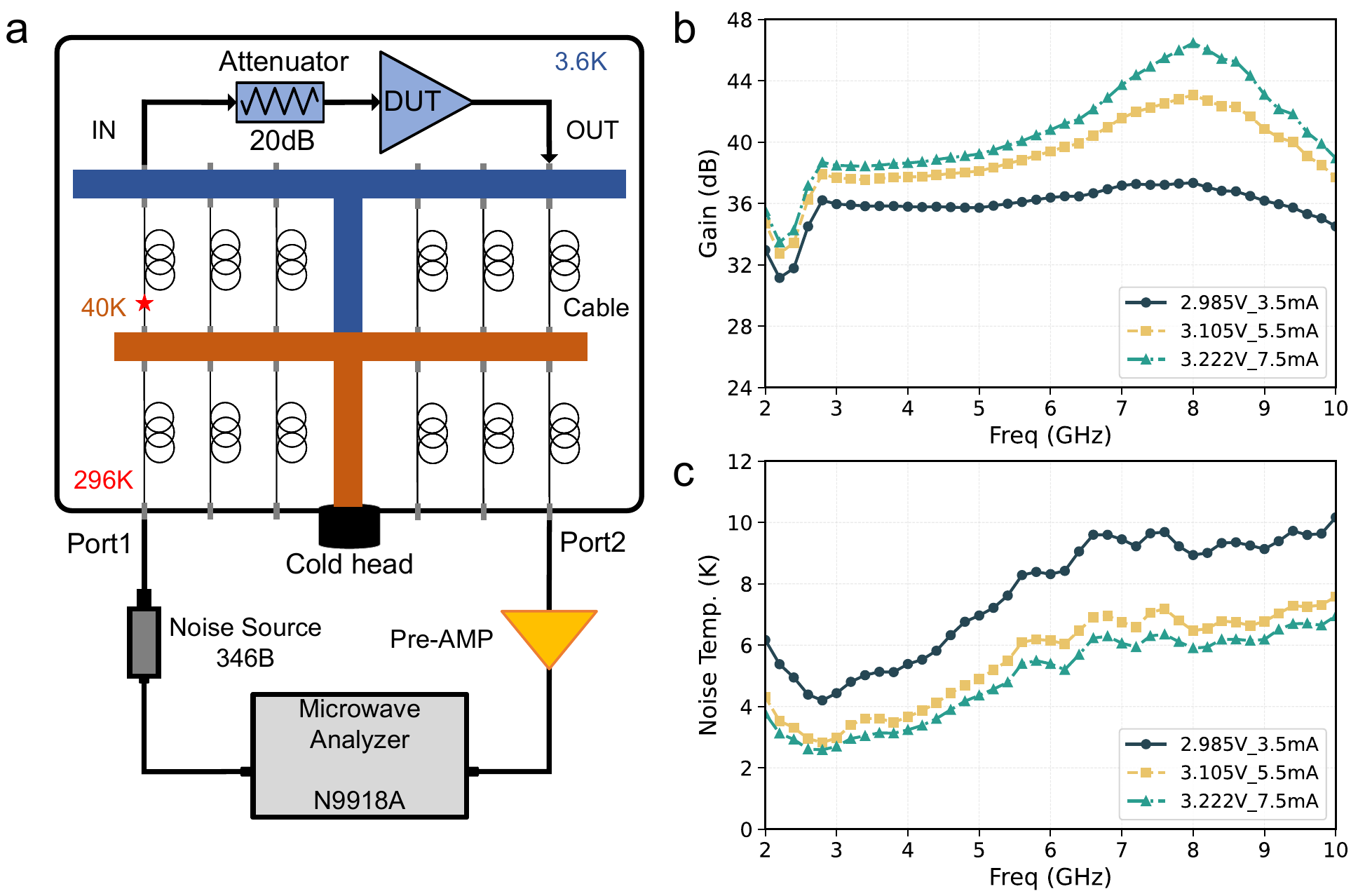}
    \caption{(a) Schematic of the measurement setup for benchmarking the cryo-LNA at 3.6~K. The device under test (DUT) is connected between an input attenuator (20~dB) and an output port, with thermal stages maintained at 296~K, 40~K, and 3.6~K. A noise source (346B) and a microwave analyzer (Keysight FieldFox N9918A) are used to measure gain and noise temperature. (b) Measured gain of the cryo-LNA across the 2–10~GHz frequency range under three different bias conditions. The gain exceeds 32~dB across the entire range and peaks at over 44~dB under the highest bias condition. (c) Measured noise temperature of the cryo-LNA, remaining below 10~K across the full range and dropping to less than 6~K in the 5–7 GHz range for the highest bias condition. }
    \label{fig:4}
\end{figure*}

The cryo-LNA block was initially tested at an ambient temperature ($T_a$) of 300~K, with the supply voltage ($V_{dd}$) varied from 3.4~V to 4.0~V. A Rohde \& Schwarz vector network analyzer (ZNB 20) was used to measure the scattering parameters (S-parameters), while a Keysight FieldFox N9918A microwave analyzer with a noise figure measurement option was employed to evaluate noise performance.

The gain performance of the designed cryo-LNA is shown in Fig.~\ref{fig:3}(a). Across the frequency range from 2~GHz to 10~GHz, the gain increases with the supply voltage, achieving a maximum gain exceeding 40~dB under the highest bias condition. In the frequency range of 5 to 7~GHz, which is relevant for superconducting qubit readout, the amplifier achieves a gain exceeding 33~dB even under the lowest bias condition. These results demonstrate that the designed cryo-LNA achieves high gain under various operating conditions. 
Figure.~\ref{fig:3}(b) presents the noise performance of the designed cryo-LNA. The noise figure (NF) remains below 1.5~dB across the entire frequency range for all bias conditions. Specially, in the 5 to 7~GHz range, the cryo-LNA achieves a minimum noise figure of less than 0.9~dB, which is crucial for enabling high-fidelity readout of superconducting qubits. 
Figure.~\ref{fig:3}(c) and Figure.~\ref{fig:3}(d) show the input and output return losses of the deigned cryo-LNA. The input return loss ($S_{11}$) remains below $-10$~dB in the 4 to 8~GHz range, indicating good input impedance matching across the range. Similarly, the output return loss ($S_{22}$) remains below $-15$~dB in the same frequency range, demonstrating effective output impedance matching under varying bias conditions.   

Characterizing the noise performance of amplifiers at cryogenic temperatures is challenging due to the need for a precise reference noise level at the amplifier's input to accurately measure absolute noise~\cite{Simbierowicz2021,Malnou2024}. After verifying the performance at 300~K, additional efforts were devoted to characterizing the LNA at cryogenic temperature ($T_a = 3.6~\rm{K}$), following a setup similar to that in~\cite{Peng2023}. As shown in Fig.~\ref{fig:4}, the LNA modules were installed inside a 4 Kelvin cryostat, and $T_e$ was measured using the cold-attenuator method. In this setup, a cold/hot noise source (Keysight 346B) operating at 300~K provided the noise reference for measurement. A 20~dB coaxial attenuator was connected at the LNA input port to ensure thermal isolation. A pre-amplifier operating at 300~K provided sufficient gain for noise power measurement, which was performed using a microwave analyzer (Keysight FieldFox N9918A). The noise source provided different noise powers corresponding to the physical temperatures in its ON and OFF states. The analyzer then read these power levels and performed the necessary calculations. Additionally, the insertion loss between port 1 and port 2 was measured by removing the LNA in a similar setup. These measurements were used to calibrate cable loss and de-embed the added noise contributions from fixtures, including cables, attenuators, and adapters. One significant source of noise at the front stage is the coaxial cable running from 296~K to 3.6~K, with the cable temperature at the midpoint estimated to be approximately 40~K. These adjustments enabled the accurate calculation of $T_e$ from the measured data~\cite{Sheldon2021}.

\begin{figure*}[t]
    \centering
    \includegraphics[width=0.8\textwidth]{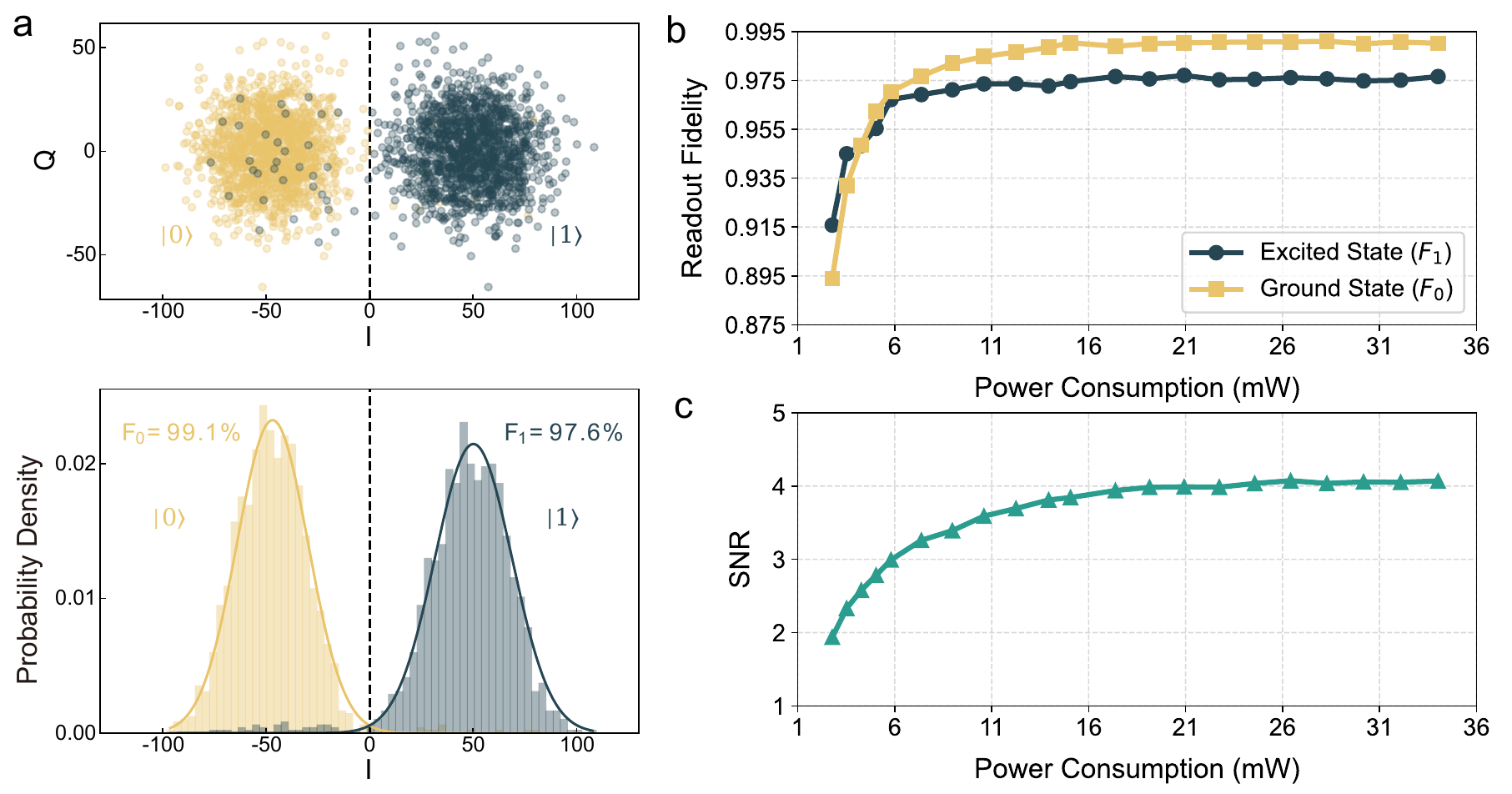}
    \caption{(a) The achieved I/Q scatter and histogram plots using the designed LNA for superconducting qubit measurement. (b) Measured readout fidelity of correctly identifying the ground state and the excited state varing with the DC power consumption of the designed cryo-LNA. The readout fidelity exceeds 95\% at power levels as low as 6~mW and stabilize above 99\% and 97.5\%, respectively, for power levels exceeding 16~mW. (c) Measured SNR of the superconducting qubit readout signal as it varies with the DC power consumption of the cryo-LNA. The SNR increases steadily with power consumption, reaching approximately 3.5 at 10mW and peaking at 4 under the maximum power condition (31~mW).}
    \label{fig:5}
\end{figure*}

The gain of the cryo-LNA was measured across the 2–10 GHz frequency range under three different bias conditions. The results, shown in Fig.~\ref{fig:4}(b), indicate that the gain increases steadily with higher supply voltage and current. Across the entire frequency range, the gain exceeds 32~dB under all conditions and reaches over 44~dB for the maximum bias setting. In the citical frequency range of 5-7~GHz, which is relevant for the superconducting qubit readout, the amplifier consistently delivers a gain of at least 36~dB,demonstrating it suitability for quantum computing applications. 
Figure.~\ref{fig:4}(c) shows the noise temperature of the cryo-LNA measured across the same frequency range and bias conditions. The noise temperature remains below 10~K throughout the entire frequency range for all conditions. Notably, under the highest bias condition, the noise temperature drops below 6~K in the 5-7~GHz range. This ultra-low noise performance is critical for high-fidelity readout of superconducting qubits, where minimizing noise is essential for preserving quantum information.

These results demonstrate that the designed cryo-LNA achieves high gain and ultra-low noise temperature at cryogenic temperatures, validating its potential for superconducting qubit readout and other cryogenic applications requiring low-noise amplification.

\section{BENCHMARK WITH SUPERCONDUCTING QUBITS}
\label{5}

To benchmark the cryo-LNA performance, we incorporate the cryo-LNA in the measurement of a qubit, sharing a configuration similar to that depicted in Fig.~\ref{fig:1}(a). The LNA module was positioned on a 4~K plate within the BlueFors DR. Notably, we did not employ a Josephson Parametric Amplifier (JPA), making the noise performance of the entire chain reliant on the LNA's noise performance~\cite{Aumentado2020}. The signal generated by the Digital to Analog Converter (DAC) is up-converted to microwave frequency range with an IQ mixer, then transmitted to the quantum chip through the attenuators and low-pass filters (LPF) from 300~K to mixing chamber of $\sim10~\rm{mK}$. It is then amplified and transmitted to the Analog-to-Digital Converter (ADC) for processing~\cite{Zhang2024}.

The results of I/Q demodulation for the qubit state are illustrated in Fig.~\ref{fig:5}(a). The ground state ($|0\rangle$) and excited state ($|1\rangle$) are well-separated in the quadrature space, clearly demonstrating distinguishability enabled by the cryo-LNA's high gain and low noise performance.
The corresponding histograms display the probability density distributions for both states. The ground state achieves a fidelity of $F_0 = 99.1\%$, while the excited state achieves $F_1 = 97.6\%$, reflecting high measurement accuracy. The measurement results, which illustrate the variations in readout fidelity with the DC power consumption of the cryo-LNA, are shown in Fig.~\ref{fig:5}(b). The readout fidelity improves as power increases and stabilize above $F_{0} = 0.99$ and $F_{1} = 0.975$ when the power exceeds 16~mW. Importantly, even at lower power levels of around 6~mW, the readout fidelity remains high, exceeding 95\%, demonstrating that reliable qubit state discrimination can be achieved at relatively low power consumption. Figure.~\ref{fig:5}(c) illustrates the variations in the SNR of the qubit readout signal as the DC power consumption increases. The SNR grows steadily with power and reaches a peak value of approximately 4 at 31~mW. However, it is noteworthy that an SNR of around 3.5, sufficient for accurate qubit state readout, is achieved at a much lower power consumption of approximately 10~mW. 

Together, these results emphasize the cryo-LNA’s capability to provide reliable qubit readout performance while maintaining low power consumption, which is critical for large-scale quantum computing systems with stringent power constraints.

\section{CONCLUSION}
\label{6}

In this paper, we successfully developed an MMIC cryo-LNA utilizing a 150~nm GaAs pHEMT process. The LNA incorporates a current multiplexing structure to minimize power consumption and features a negative feedback loop to ensure broadband stability. It has a compact chip size of only 1.85~mm × 1.2~mm. Within the frequency range of 4 to 8~GHz, the LNA demonstrates an average gain of 38~dB and a minimum equivalent noise temperature of 5~K when operating at 3.6~K.
We benchmarked the LNA's performance with a superconducting qubit, achieving an average single-shot dispersive readout fidelity of 98.3\% without assistance from a quantum-limited parametric amplifier. Notably, even at a low DC power consumption of just 6~mW, the LNA provides reliable readout fidelity, demonstrating its capability to balance energy efficiency and high performance. The development of this GaAs MMIC cryo-LNA, suitable for cost-effective and large-volume manufacturing, diversifies the technical support available for large-scale quantum applications.

\section{ACKNOWLEDGMENTS}
This work was supported by the Science, Technology and Innovation Commission of Shenzhen Municipality (KQTD20210811090049034), the Innovation Program for Quantum Science and Technology (2021ZD0301703), Shenzhen Science and Technology Program (Grant No. RCBS20231211090815032, RCBS20231211090824040).

\bibliography{Ref1}
\end{document}